# Design and Test of Small Mirror Supports for Harsh Environments

Ruby Huie[1], Austin Mears[1], Manny Montoya[1], Dan Vargas[1], Grant West[1], Daniel Hofstadter[1] and Ewan S. Douglas[1]
[1]Center for Astronomical Adaptive Optics, 933 N Cherry Ave, Tucson, AZ 85719; Dept. of Astronomy

## ABSTRACT

As wavefront quality demands tighten on space systems for applications such as astronomy and laser communication, mounting small optics such that the wavefront is undisturbed, positioning is adjustable and the design is producible, while surviving harsh space environments, is a continuing challenge. We designed multiple candidate flexure mounts to support small optics (up to 50 mm diameter, and over 100 grams) to survive the qualification and acceptance tests of small spacecraft and units as defined in ISO 19683 and a mounting structure which is adjustable in decenter [+/-0.5mm], tip/tilt +/-0.5deg, and piston [+/-0.25mm]. We will present design details along with measurements showing less than approximately lambda/10 wavefront contribution from the optic bonding process, along with thermal and multi-axis vibration test data showing the mounted optics survived the acceptance testing loads and are suitable for operation in a wide range of harsh environments.

**Keywords:** wavefront quality, small optics, adjustable mounting structure, space systems

## 1. INTRODUCTION

As the demands for high-precision optical systems in space applications steadily increase in the fields of astronomy and laser communication, as have the challenges associated with achieving this. High quality wavefronts are crucial in these applications, and thus necessitate mounting solutions that ensure the wavefront is undisturbed, their position is adjustable, and the design is producible, all whilst being able to survive the harsh conditions of space.

One of the challenges associated with the use of small optics in space is the ability to create mounts that are capable of withstanding the extreme mechanical and thermal stresses endured in space without endangering the optical quality. Traditional mounting solutions often fail to provide the necessary stability and adjustability which leaves room for degradation in the wavefront quality. This issue is then further compounded by the stringent requirements that space environments have for positional accuracy and structural integrity.

To address this challenge, we designed and mounted multiple candidate flexure mounts that are able to support small optics (up to 50 mm diameter, and over 100 grams). These mounts were engineered to survive the qualification and acceptance tests of small spacecrafts as defined in ISO 19683. Our mounts offer precise adjustability in decenter (± .5mm), tip/tilt (± .5 degrees), and piston (± .25mm) which ensure accurate alignment and position of the optics.

This paper presents the design details and experimental validation of these flexure mounts. To validate the mounts, comprehensive static and dynamic structural analyses were performed as well as extensive thermal testing. The results were then verified using interferometry and faro arm analysis. The subsequent results demonstrate that our mounts are capable of enduring the extreme conditions that can be expected during launch and operation while maintaining minimal wavefront degradation.

The successful validation of these mounts carries significant implications for the future of high precision optical systems in space applications. They have the potential to reduce the cost and time often associated with developing new mounting solutions without impacting the accuracy and efficiency of astronomical observations and laser communications. Their validated resilience under a multitude of harsh conditions will help to drive the progress of space exploration and technology. [16]

## 2. RELATED WORK

The development and implementation of small mirror supports for space systems have been extensively studied with significant contributions focusing on the design, thermal stability, and structural integrity under space conditions. It is through this prior existence of literature and standards that we were able to shape our understanding and approach for creating our own small mirror supports.

Standards such as ISO 19683 and the General Environmental Verification Standard (GEVS) are crucial in helping to define the qualification and acceptance criteria for optical mounts. ISO 19683 outlines the environmental testing requirements for small spacecraft. This ensures that all components are able to withstand the conditions that will be encountered during both launch and operation. This standard incorporates a comprehensive range of tests, including thermal cycling, vibration, and shock testing. Adhering to this standard was critical to the success of our small mirror supports. GEVS also helped to provide us with guidelines for the environmental verification that NASA's spaceflight hardware utilizes. GEVS outlines the methodologies required for conducting tests to simulate the stresses and conditions experienced during launch and operation. By following these standards, we were able to ensure that our optical mounts could meet the stringent requirements for space missions, thereby minimizing the overall risk of failure. [12][15]

This literature highlights the challenges often associated with developing small mirror supports for space applications. By building on these foundational works and adhering to established standards, our own research will help to contribute to further advancing high-precision optical systems.

## 3. METHODOLOGY

This section details the methodology employed to design our candidate flexure mounts and evaluate their subsequent performance under expected structural and thermal environments. The goal was to design a flexure mount that could be precisely manufactured, precisely assembled, and survive hardware life cycle details as defined in ISO19683 and GEVS, specifically the dynamic and thermal environments. These two standards implement rigorous testing to validate hardware performance required for space missions. As such, experimental and inspection level testing was performed to address and validate hardware performance for these defined conditions.

**3.1 Mechanical Design**

The mechanical design is a crucial aspect towards ensuring that the mounts are durable and will satisfy performance criteria. Design considerations looked at previous space missions, publications and derived concepts. The first design looked at was an existing Osiris-Rex design. This configuration was for an Osiris-Rex OCAMS filter that consisted of a singular filter holder with built in radial flexures. The simplistic design geometry meant that standard machine operations would be able to produce it. The second design was created with the goal of being able to control the placement and height of the optic. The design offered injection holes on the sides of the flexures to allow for the application of adhesives. Iterating on the second design, a third radial design was established. The radial flexure design allowed for symmetric distortion while also reducing the likelihood for thin wall flexures which can be produced from a single rod stock. This design allowed precise control over the placement of the flexures relative to the optic's center of gravity. An additional flexure design was considered for the fourth design. The design allowed for more precise control of the locations of the flexure arms using GD&T vs relying on machining capability of built-in flexures. This design also allowed it to act as a universal mount to be able to accommodate optics and flexures of variable sizes. These designs can be seen below in *Figure 1*.

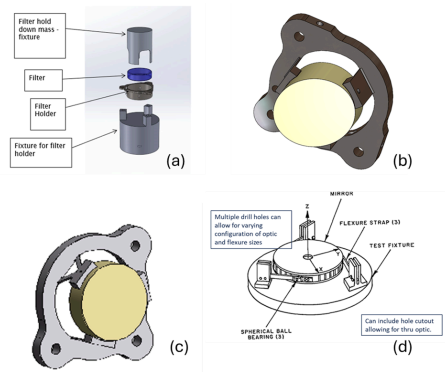

*Figure 1: Preliminary Mechanical Design Models (a) First Design (b) Second Design (c) Third Design (d) Fourth Design*

Subsequent iterations led to an amalgamation of the radial design and flexure arm design. This design combines the radial flexure's ability to allow for symmetric distortion with the flexure arms and strategically placed injection holes for adhesive application.. The final assembly of the chosen best candidate includes the radial cell integrated with a mounting fixture. The mounting fixture consists of a stage, and decenter stage. The complete assembly is depicted below in *Figure 2*. The mount is also available for reuse at [Zenodo](Zenodo).

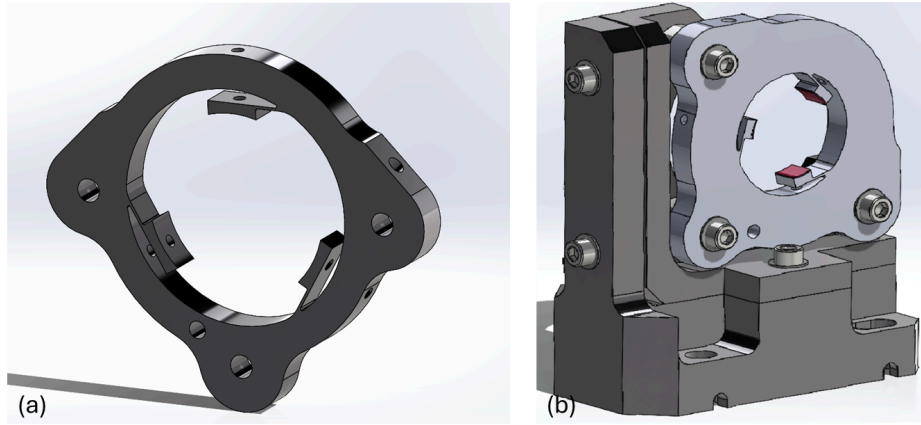

*Figure 2: (a) Optic Cell (b) Optic Assembly*

### 3.2 Material Selection

The selection of materials for the optical mounts was a critical step towards ensuring the components overall performance and reliability. As such, a comprehensive material selection process was undertaken. This process involved detailed simulations and analyses utilizing ANSYS software to evaluate the mechanical, thermal, and adhesive properties of various material combinations.

To begin the material selection process, the criteria for material selection had to be defined. The harsh environment of launch and operation necessitated stringent requirements to ensure that the optical mounts maintained their reliability and precision. The primary considerations in material selection were the mechanical strength, thermal stability, and compatibility with the adhesives used in assembly.

Based on the established criteria, a trade study was conducted to identify the most suitable materials and adhesives for the optical mount. The established candidate materials are shown in *Table 1*. These candidate materials were initially selected based on their hardware properties at temperature and their ability to meet the performance requirements of the optic mount.

| Material Properties | Mount Material | | | Adhesive Material | | | | Optic Material |
|---|---|---|---|---|---|---|---|---|
| Description | Aluminum 6061-T6 [1] | Invar36® [2] [3] | Ti-6Al-4V [4] | 3M™ 2216 B/A Gray [5] | Milbond [6] | Urethane [7] (Appli-Thane®7300) | RTV 566 [8] [9] | Zerodur® [10] |
| Density (ρ) | 0.098 lb/in^3 | 0.291 lb/in^3 | 0.160 lb/in^3 | 0.049 lb/in^3 | - | 0.101 lb/in^3 | 0.055 lb/in^3 | 0.091 lb/in^3 |
| Shear Modulus (G) | 3800 ksi | 8122 ksi | 6200 ksi | 260 ksi @-30°C | 80 ksi @-30C | 3610 psi @RT | 92.1 psi @RT | - |
| Lap Shear @RT | - | - | - | 3200 psi | 2099 psi | 500 psi | 465 psi | - |
| Coefficient of Thermal Expansion (CTE) (α) | +185°F (+85°C) 12.95E-6/°F (23.3E-6/°C) +72°F (+22°C) 12.65E-6/°F (22.7E-6/°C) -22°F (-30°C) 12.3E-6/°F (22.1E-6/°C) | +185°F (+85°C) 0.8E-6/°F (1.5E-6/°C) +72°F (+22°C) 0.8E-6/°F (1.5E-6/°C) -22°F (-30°C) 0.7E-6/°F (1.3E-6/°C) | +185°F (+85°C) 5E-6/°F (9E-6/°C) +72°F (+22°C) 4.9E-6/°F (8.82E-6/°C) -22°F (-30°C) 4.8E-6/°F (8.64E-6/°C) | +185°F (+85°C) (134E-6/°C) +72°F (+22°C) (102E-6/°C) -22°F (-30°C) 39.4E-6° F (70.9E-6/°C) | +185°F (+85°C) (72E-6/°C) +72°F (+22°C) (72E-6/°C) -22°F (-30°C) (62E-6/°C) | +185°F (+85°C) (75E-6/°C) +72°F (+22°C) (75E-6/°C) -22°F (-30°C) (75E-6/°C) | +185°F (+85°C) (233E-6/°C) +72°F (+22°C) (233E-6/°C) -22°F (-30°C) (233E-6/°C) | +185°F (+85°C) (0.006E-6/°C) +72°F (+22°C) (0.006E-6/°C) -22°F (-30°C) (0.006E-6/°C) |

*Table 1. Candidate Material Mechanical Performance*

Daly and Hawk [11] helped provide the down-selection criteria for identifying an acceptable optical adhesive that could meet performance standards. The down-selection criteria was established utilizing the two equations provided for identifying the resultant shear stress (τ) of the adhesive that occurs during temperature change (ΔT) multiplied by the adhesive shear modulus (G), the bond diameter (a), the bond thickness (t), and the difference between the coefficient of thermal expansion (CTE) of the optic ($α_1$) and the CTE of the optic mount ($α_2$) as shown in eq.1

$$\tau = \frac{Ga}{2t}(α_1 - α_2)\Delta T \qquad \text{eq.1}$$

The equation for identifying the minimum required bond area is shown in eq.2 where the max load is defined as the product of the weight of the optic or 0.04lb for the weight of a small 1in optic, the safety factor for bonds to glass or 2.6SF [12], and the maximum equivalent shock response spectra or 100G for the optic [12]. This methodology can be considered a conservative approach for identifying the minimum bond area required for an unknown, system level, shock response for a spacecraft that has a complex airframe. Though the shock load is high, it is a conservative estimate for the hardware to be placed anywhere within the spacecraft.

$$Minimum\ bond\ area\ =\ Max\ Load\ /\ Adhesive\ Shear\ Strength \qquad \text{eq.2}$$

Using the two equations above, *Table 2* shows the resultant induced thermal shear load and *Table 3* shows the minimum bond area required for the candidate adhesive material. For equation 1, the temperature delta was established as the temperature that would cause the largest gradient or the delta between the hot condition (+85C) and the room temperature condition (+22C). The bond diameter was chosen as 0.5in and the bond thickness was selected as 0.010in for demonstration purposes. Note that adhesive manufacturer's typically provide a recommended bond line thickness in the material's technical data sheets. As shown in *Table 2*, the best outright performer for inducing a low thermal shear stress on the optic was the combination of Invar36 and Zerodur that kept all of the resultant shear stress loads to less than 1000psi. This would comply with standard industry practice for keeping low residual stress in the optical adhesive [13] [14]. Due to the two material's low CTE values, they are more capable of responding to temperature at nearly the same rate in comparison to the aluminum or titanium materials. As shown in *Table 3*, the material requiring the least amount of application area to achieve the same strength response under load was identified as the 3M 2216 B/A Gray epoxy adhesive. Therefore, the material that would meet all the objectives of the environmental requirements would be the combination of the Invar36 mount material, a Zerodur optic and the 3M 2216 B/A Gray epoxy adhesive.

| Assembly Configuration | | Mount: Aluminum 6061-T6 Optic: Zerodur® | Mount: Invar36® Optic: Zerodur® | Mount: Ti-6Al-4V Optic: Zerodur® |
|---|---|---|---|---|
| Adhesive | 3M™ 2216 B/A Gray | 9538.89psi | 611.79psi | 3683.04psi |
| | Milbond | 2935.04psi | 188.24psi | 1133.24psi |
| | Urethane | 132.44psi | 8.49psi | 51.14psi |
| | RTV 566 | 3.38psi | 0.22psi | 1.30psi |

*Table 2. Resultant Induced Thermal Shear Load Comparison for each Assembly Configuration*

| Material | Adhesive Strength (psi) | Resulting Total Minimum Bond Area (in^2) |
|---|---|---|
| RTV-566 | 465 | 0.022 |
| Urethane Epoxy | 500 | 0.021 |
| Milbond | 2099 | 0.005 |
| 3M™ 2216 B/A Gray | 3200 | 0.003 |

*Table 3. Resultant Minimum Bond Area Required for Small Optic under 100G Load*

To confirm the material selection and increase confidence in the design, this combination was further evaluated through detailed simulations and testing in ANSYS. In the ANSYS simulation environment, the optic cell and optic mount were modeled with the appropriate geometric details and boundary conditions, as shown in *Figure 3*.

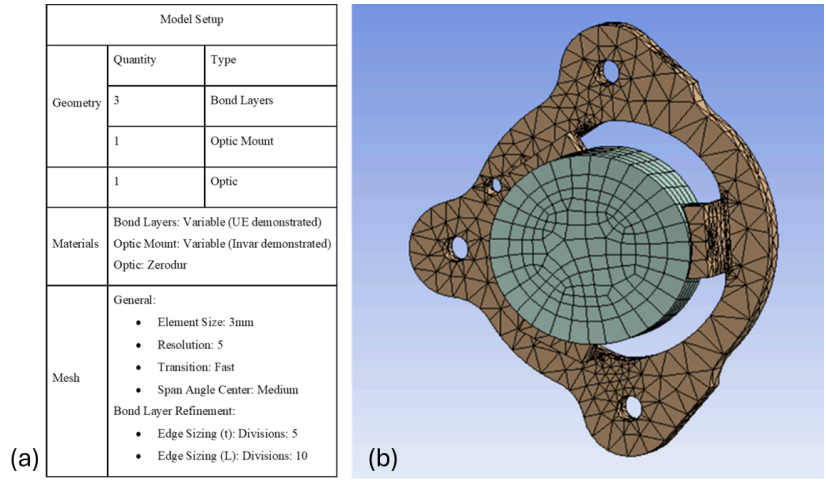

*Figure 3: (a) ANSYS Simulation Model Setup (b) ANSYS Simulation Model*

Mechanical loads simulating launch conditions were applied to the model to understand the mechanical behavior of the chosen design under dynamic conditions. When the Maximum Permissible Exposure (MPE) +3dB random vibration profile was applied to the optic cell, an equivalent stress of approximately 33MPa (4.8ksi) was observed at the flexure base. The minimum yield strength of Invar36 is approximately 33 ksi, indicating that the chosen material can withstand the applied loads without yielding. The axes response and ANSYS model can be seen in *Figure 4*.

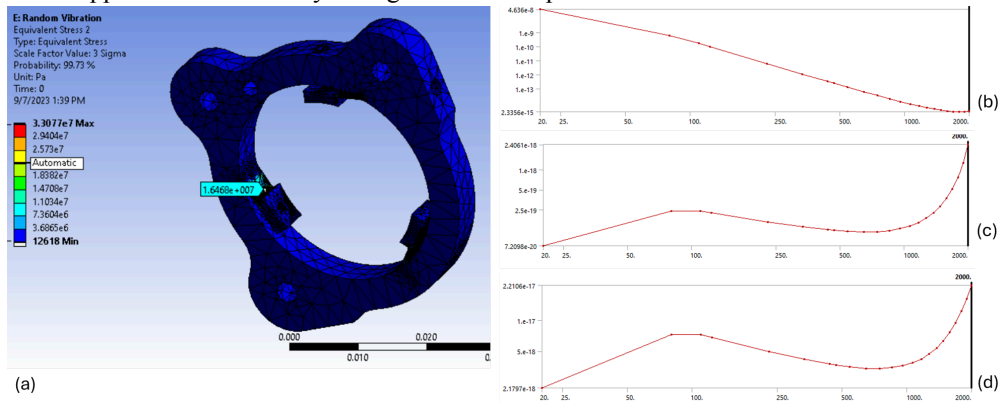

*Figure 4: (a) Optic Cell ANSYS Model (b) X-Axis Response from a Y-Axis Random Vibration Input (c) Y-Axis Response from a Y-Axis Random Vibration Input (d) X-Axis Response from a Y-Axis Random Vibration Input*

Using the modal analysis tool in ANSYS on the chosen cell design with the optic installed, resulted in the first mode being activated at approximately 2.3kHz, as shown in *Figure 5*.

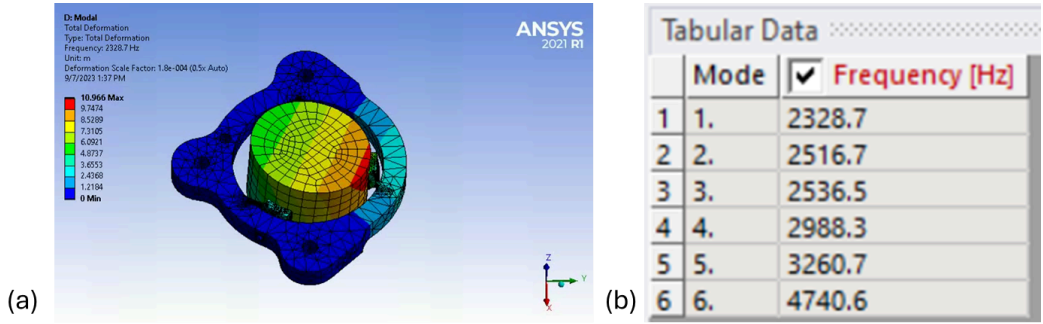

*Figure 5: (a) Optic Cell ANSYS Model (b) Modal Response of Optic Cell*

With the optic cell installed into the optic assembly, the complete assembly's first mode shifted to approximately 1.2kHz. Applying the MPE +3dB random profile to the whole optic assembly resulted in an equivalent stress of approximately 25MPa (3.6ksi) at the flexure base. This stress is significantly lower than the yield strength of Invar36, reinforcing that the complete assembly can endure the applied random vibrations without permanent deformation. It is important to note that the equivalent stress dropped from 33Mpa in the isolated optic cell to 25Mpa in the assembled state. This reduction in stress highlights the stress mitigation that integrating the optic cell into the larger optic assembly provides. The observed lower stress response can be attributed to the increased mass of the complete assembly, which absorbs more of the applied energy, resulting in a lower stress response on the flexure arms. The axes responses from the random vibration input seen below in *Figure 6* illustrate the dynamic behavior of the system and further corroborate that the chosen materials and design provide sufficient robustness.

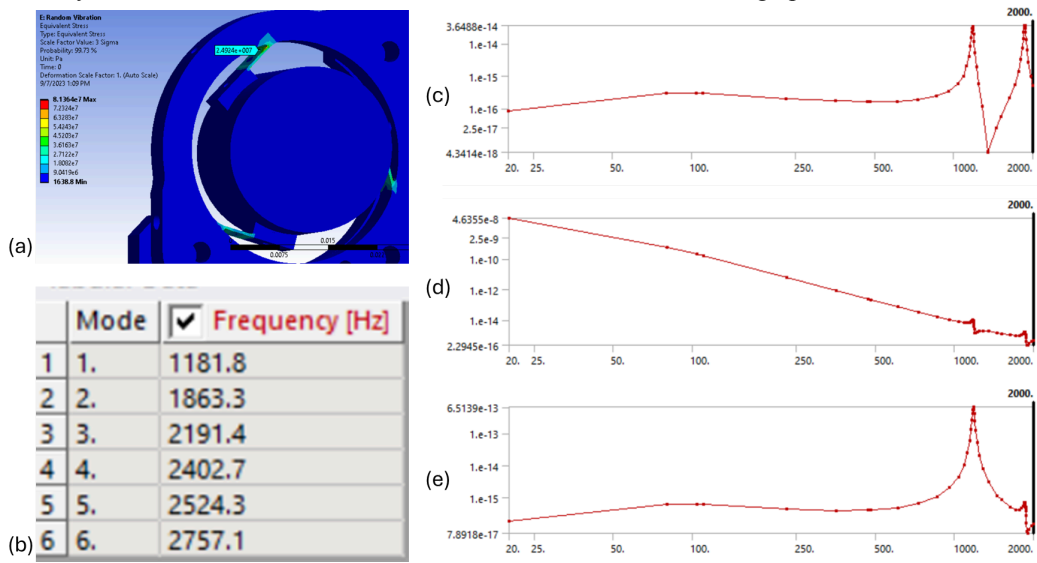

*Figure 6: (a) Random Vibration Response of Flexure Arms in Optic Assembly (b) Modal Response of Optic Assembly (c) X-Axis Response from a Y-Axis Random Vibration Input (d) Y-Axis Response from a Y-Axis Random Vibration Input (e) Z-Axis Response from a Y-Axis Random Vibration Input*

### 3.3 Experimental Test Setup

This section outlines the experimental setup and procedures used to evaluate the performance of our candidate flexure mounts under simulated operational conditions. The experimental tests were designed to replicate the dynamic and thermal environments as specified in ISO 19683 and GEVS standards. Through a series of controlled tests, we aimed to validate the structural integrity, thermal resilience, and overall reliability of the flexure mounts.

### 3.3.1 Optic Cell Assembly-Dynamic

The optic cell assembly was characterized by performing sine-sweep and random vibration testing on a vibration shaker at Hofstadter Analytical Services in Tucson, Arizona. As shown in *Figure 7*, an optic cell assembly is placed on a vibration adapter plate that allows for the test unit to be tested in three (3) orthogonal directions without having to remove the test unit from the adapter. Five (5) total accelerometers would be used for testing. Three (3) single axis accelerometers were placed at approximate orthogonal locations on the test unit to measure the x, y and z directional responses. The remaining two (2) single axis accelerometers were used for shaker control and monitoring input levels. The test equipment used during testing can be seen below in *Table 4*.

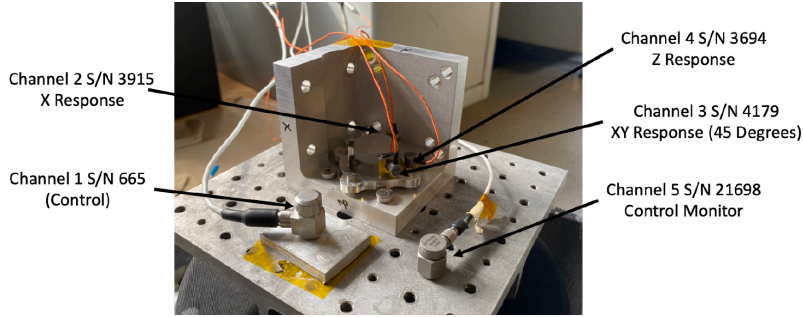

*Figure 7: Dynamic Testing Setup – Optic Cell Assembly*

| Item | Manufacturer | Model | Serial Number | Calibration |
|---|---|---|---|---|
| Test Article | UASO | N/A | 1-5 | N/A |
| Shaker | Vibration Test Systems | VG 150A 6 | 142 | N/A |
| Controller | Dynamic Solutions | DVC8/8 | BW618 | 9/1/22 |
| Accelerometer | Dytran | 3006A | 665 | 6/8/22 |
| Accelerometer | Dytran | 3032A | 3915 | 6/8/22 |
| Accelerometer | Dytran | 3032A | 4179 | 6/8/22 |
| Accelerometer | Dytran | 3032A | 3694 | 6/8/22 |
| Accelerometer | Dytran | 3005D4 | 21698 | 6/8/22 |

*Table 4: Testing Equipment*

For testing, sine-sweeps occurred pre- and post- random vibration tests to ensure the test unit had no premature failure that would otherwise not be visible to the unaided eye. Sine sweeps were performed at a 1.0g level, 18 Hz to 3500 Hz, swept at 3.0 octaves per minute as defined in *Figure 8a* and *Figure 8b*. ISO19683 specifies a higher-level sine sweep over a narrower frequency range (5 Hz to 100 Hz) to simulate an 8.4G (minimum) quasi-static load. This profile was modified to to serve as a modal survey over a wider frequency range, while still serving to identify any damage not readily apparent. At a minimum, one sine-sweep was performed after each random vibration exposure.

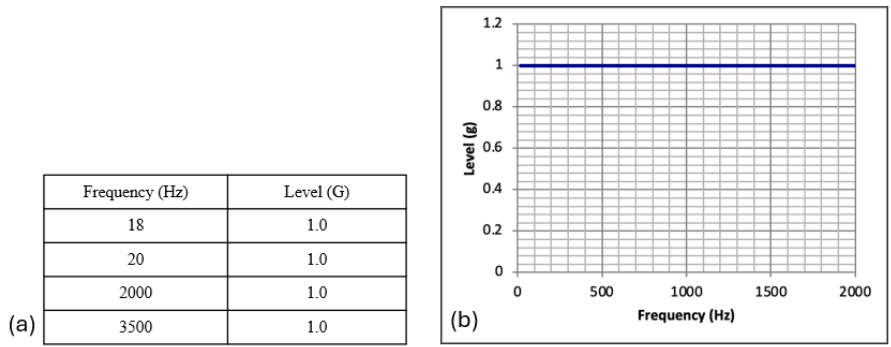

*Figure 8:(a) Sine Sweep Breakpoint Table and (b) Sine Sweep Profile*

Random vibration testing utilized the same setup as shown in *Figure 7*. To qualify the assembly, the assembly was tested to ISO19683 MPE+3dB random vibration qualification levels shown in the breakpoint table, *Figure 9a*, and the Power Spectral Density (PSD) input, *Figure 9b*. In comparison to GEVS, the ISO PSD response exceeds GEVS and would therefore qualify to both standards if hardware passed without failure. Hardware would be tested in all three (3) axes for a duration of sixty (60) seconds.

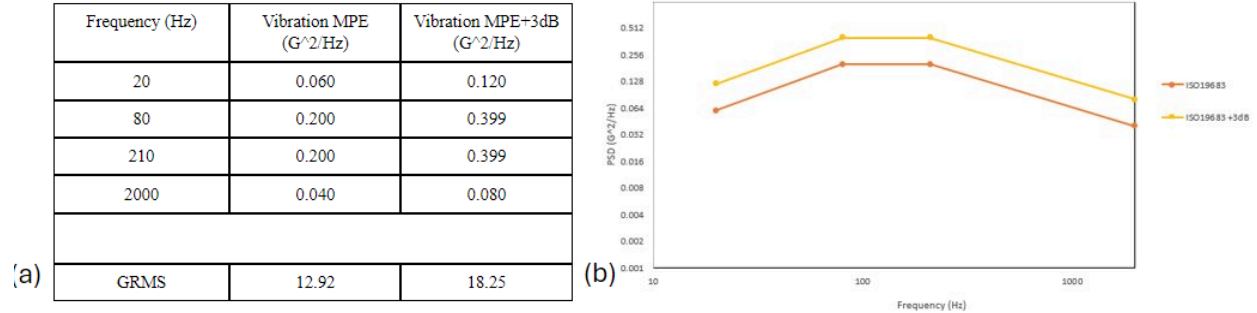

*Figure 9: (a) ISO19683 - Random Vibration Breakpoint Table and (b) ISO19683 - Random Vibration PSD Input*

### 3.3.2 Optic Cell Assembly-Mechanical Shock

The optic cell assembly was characterized for shock performance by utilizing a slide rail mechanism at Hofstadter Analytical Services in Tucson, Arizona. *Figure 10a* shows the twin-rail drop test apparatus that would impart a shock load from the select drop-heights. *Figure 10b* shows the optic cell assembly installed onto the test apparatus that would impart a shock load through the optical axis (z-direction) of the assembly. Two (2) single axis accelerometers were used for testing. One accelerometer was placed on the test apparatus mounting plate to measure in-axis acceleration. The other accelerometer was placed near the test article to measure in-axis acceleration for correlation of inputs.

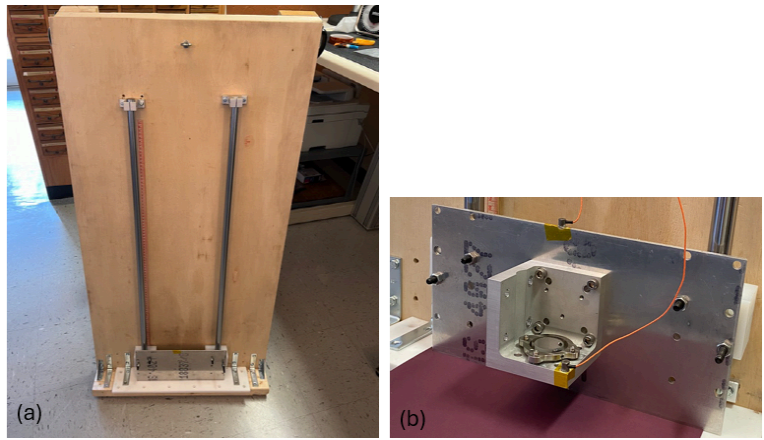

*Figure 10: (a) Twin-rail shock test apparatus and (b) test article with accelerometers placed*

Shock testing would be constructed for the assembly to test to failure and performed with a 1ms shock duration. The first target shock level was defined by referencing the ISO19683 Shock Response Spectra (SRS) as shown in Figure *Figure 11b*. Since the resonant frequency of the optic cell assembly was above 1000 Hz, the plateau level seen in *Figure 11b* was conservatively rounded up to 500G. This 500G target was initially defined to provide a margin and was progressively increased throughout testing, as seen in *Figure 11a*.

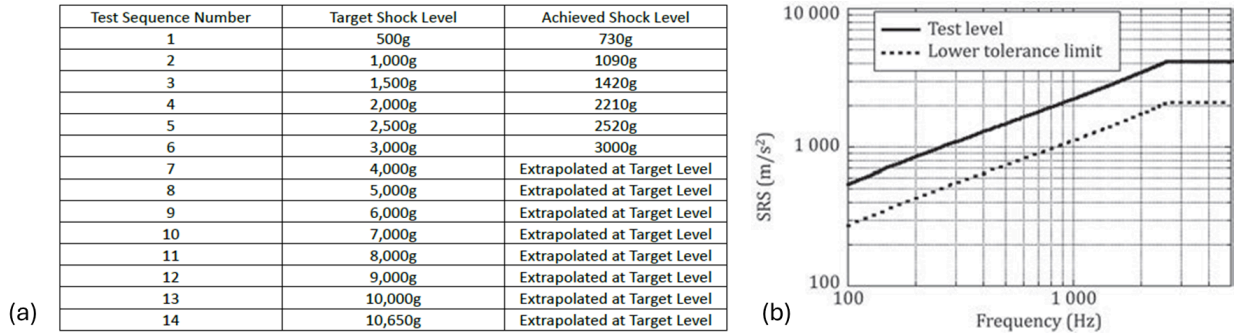

Figure 11: (a) Target Shock Levels (b) ISO19683 - Shock Response Spectra

### 3.3.3 Optic Cell Assembly and Optic Assembly-Thermal Endurance

The optic cell assembly and optic assembly were both characterized for thermal endurance by utilizing a computer controlled ESPEC thermal chamber at The University of Arizona - Steward Observatory in Tucson, Arizona. Four (4) thermocouples would be utilized for testing. As shown in *Figure 12*, three (3) thermocouples were located at the identified base, flexure and optical locations of the test article and one (1) thermocouple was placed directly adjacent to the test article to measure the ambient chamber conditions. *Figure 12* shows the testing set-up for the optic cell assembly. The testing configuration for the optic assembly can be seen in *Figure 13*.

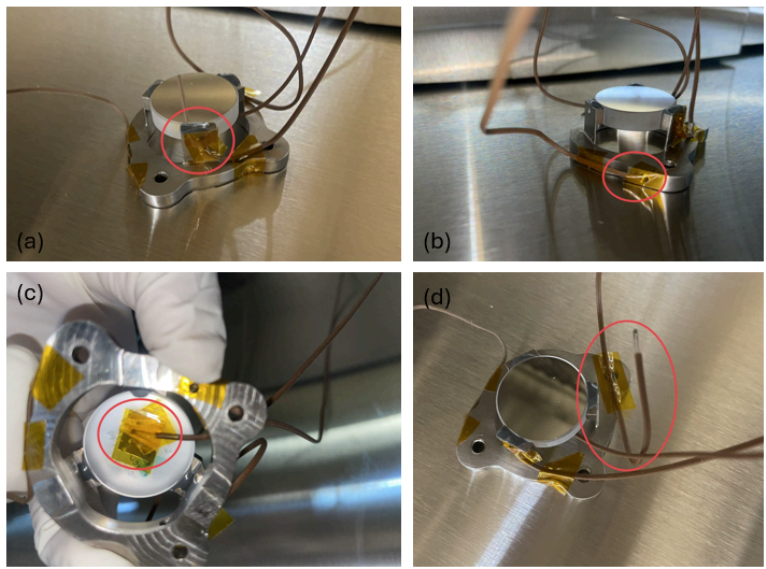

Figure 12: (a) Optic cell flexure thermocouple location, (b) optic cell base thermocouple location, (c) optic cell optic thermocouple location and (d) ambient thermocouple location

Testing was performed using the ISO19683 specification listed for thermal cycle endurance of spacecraft internal units. Temperature limits of -20C and +60C were utilized to provide margin against the specification with a temperature ramp rate of 2C/min and a soak time of 3 hrs. once the test article reached the goal temperature. The profile can be seen below in *Figure 14*.

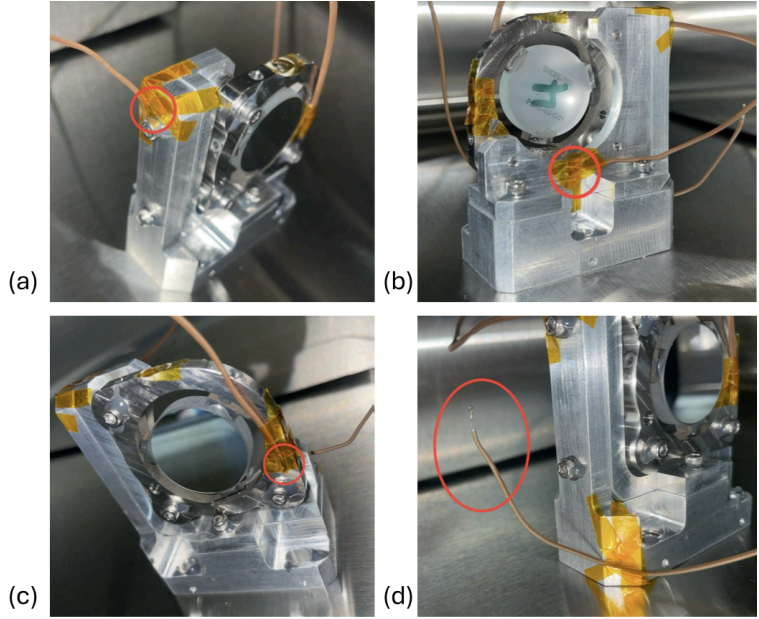

*Figure 13: (a) Optic assembly top of base thermocouple location, (b) optic assembly bottom base thermocouple location, (c) optic assembly cell base thermocouple location and (d) ambient thermocouple location*

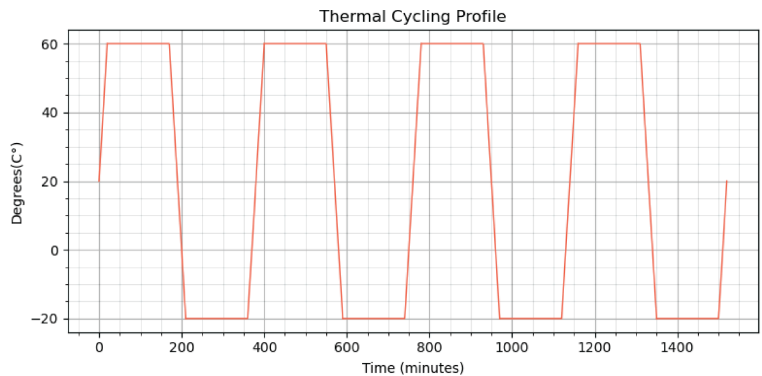

*Figure 14: Thermal Endurance Testing Profile*

The number of cycles was reduced from twenty-four (24) to four (4) to expedite the initial evaluation process. Each thermal cycling profile had a duration of 25.33 hours over four (4) cycles. Maintaining the original twenty-four (24) cycles would have extended the testing duration to an impractical length. This decision was justified on the premise that within the four cycles, the units under test would still experience the critical thermal fluctuations necessary to reveal any potential failures or material weaknesses. Subsequent extended testing would have been pursued had the preliminary tests indicated any potential issues.

### 3.3.4 Optic Assembly-Dynamic

The optic assembly was characterized by performing sine-sweep and random vibration testing on a vibration shaker at Hofstadter Analytical Services in Tucson, Arizona. As shown in *Figure 15a*, an optic assembly is placed on a vibration adapter plate that allows for the test unit to be tested in three (3) orthogonal directions without having to remove the test unit from the adapter. Four (4) total accelerometers would be used for testing. Three (3) single axis accelerometers were placed at approximate orthogonal locations on the test unit to measure the x, y and z directional responses. One (1) single axis accelerometer was placed on the shaker interface table to measure and monitor the input shaker responses. As shown in *Figure 15b*, the test coordinate system used during testing was consistent throughout all performed vibration testing. It should be noted that the z-axis is normal to the cell mounting plane.

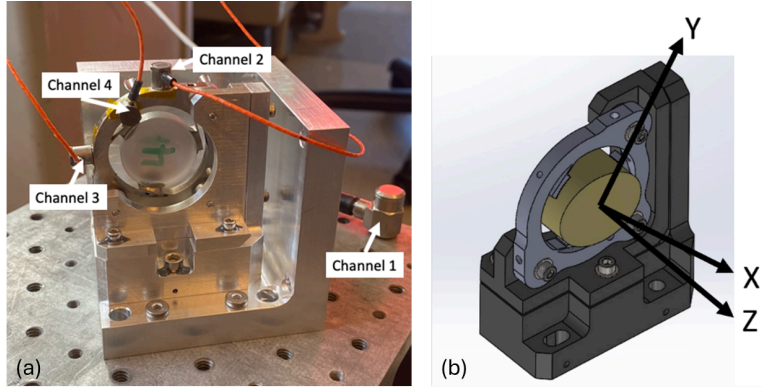

Figure 15: (a) Accelerometer Locations (Shown in Y-Axis Test Configuration) (b) Test Coordinate System

For testing, sine-sweeps occurred both prior to and after random vibration tests to ensure the test unit had not had premature failure that would otherwise not be visible to the unaided eye, similar to the optic cell assembly testing. Sine sweeps were performed at a 1.0g level, 18 Hz to 3500 Hz, swept at 3.0 octaves per minute as defined in *Figure 16a*. ISO19683 specifies a higher-level sine sweep over a narrower frequency range (5 Hz to 100 Hz) to simulate an 8.4G (minimum) quasi-static load. This profile was modified to to serve as a modal survey over a wider frequency range, while still serving to identify any damage not readily apparent. At a minimum, one sine-sweep was performed after each random vibration exposure.

Due to limitations of the vibration shaker with the increased mass of the test article, the GEVS PSD was pursued, as shown in *Figure 16b*. To accept the assembly, the assembly was tested to GEVS MPE random vibration acceptance levels. Hardware would be tested in all three (3) axes for a duration of sixty (60) seconds. This testing setup can be seen below in *Figure 17*.

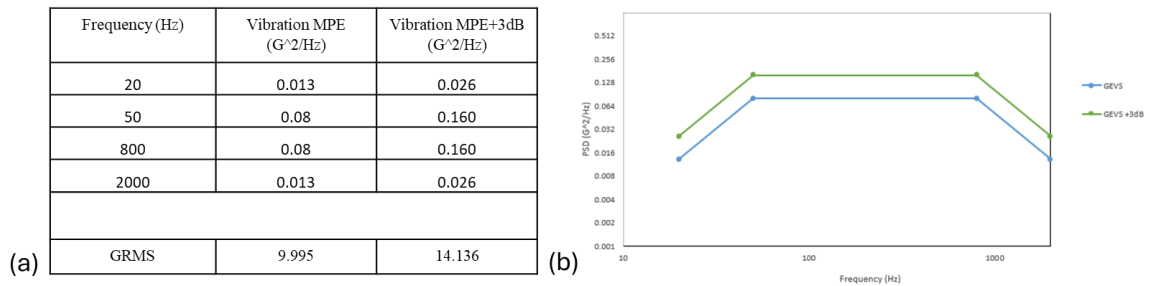

| Frequency (Hz) | Vibration MPE (G^2/Hz) | Vibration MPE+3dB (G^2/Hz) |
|---|---|---|
| 20 | 0.013 | 0.026 |
| 50 | 0.08 | 0.160 |
| 800 | 0.08 | 0.160 |
| 2000 | 0.013 | 0.026 |
| | | |
| GRMS | 9.995 | 14.136 |

Figure 16: (a) GEVS - Random Vibration Breakpoint Table and (b) GEVS - Random Vibration PSD Response

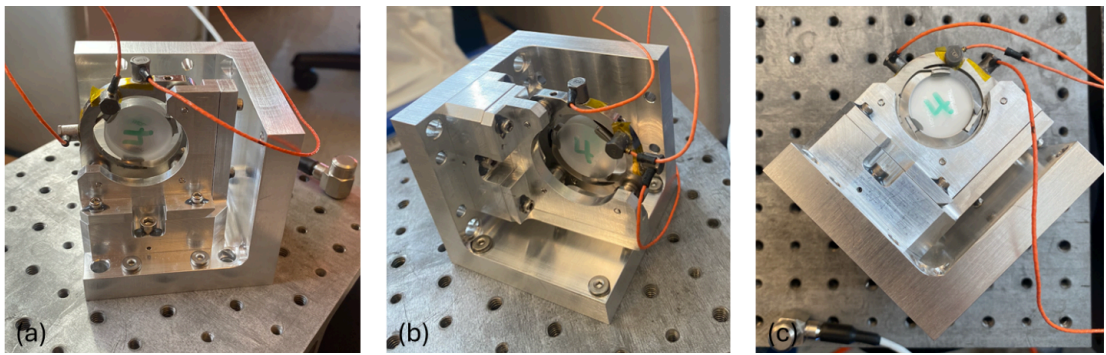

Figure 17: (a) UUT Configuration for Y-Axis Vibration (b) UUT Configuration for X-Axis Vibration (c) UUT Configuration for X-Axis Vibration

### 3.4 Post-Testing Analysis

After completing the structural and thermal testing of the candidate flexure mounts, a comprehensive post-testing analysis was conducted in order to assess the optical performance and structural integrity of each design. Interferometry and a faro arm were the tools utilized in this phase.

### 3.4.1 Interferometric Analysis

Interferometry was used to analyze the result of the structural testing as it allowed for the wavefront error of the optical mounts to be assessed. The primary tool used during this analysis was a 4D PhaseCam 6000. The interferometer was equipped with an iris diaphragm to control the aperture size and a collimator lens was utilized to ensure parallel light beams. To accurately position the optics, a 2-DOF target mount was used. The testing setup is presented in *Figure 18*.

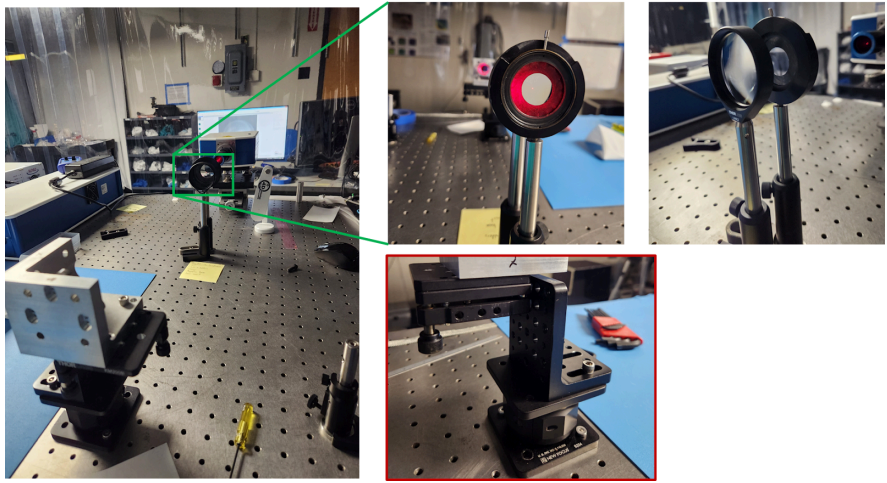

*Figure 18: Test Table Set-Up*

Each candidate flexure mount was tested individually. To determine the wavefront error induced by the structural tests, the interferometric data was analyzed post-testing with the pre-testing data as the base line. Due to resource constraints, only one mount was interferometrically tested prior to the vibe test. As such, the original pre-vibe frame was reused as the reference baseline for all subsequent post-vibe analyses. The comparison of the pre- and post-wavefront data allowed for the impacts of the structural loads on the optical mounts to be quantified.

### 3.4.2 Faro Arm Analysis

A detailed dimensional analysis of the flexure mounts was also performed following the completion of the structural and thermal tests. The analysis aimed to verify the geometric precision and dimensional stability of the mounts after exposure to the vibrational and thermal testing conditions.

A FARO Arm Coordinate Measuring Machine (CMM) with a high precision probe was used for this analysis. The FARO arm was calibrated using a certified calibration sphere in order to ensure high accuracy. To create a measurement reference frame, three perpendicular surfaces on the vibration test jig were utilized. The reference frame can be seen below in *Figure 19*.

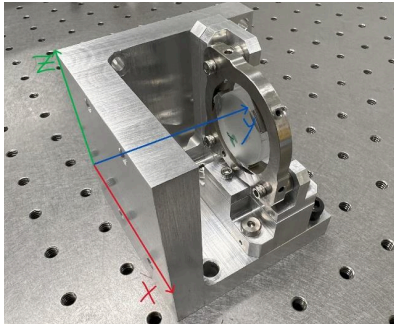

*Figure 19: Measurement Reference Frame*

Each flexure mount was positioned in a consistent orientation to ensure reproducibility. The XYZ positions of key points on the optic mount were recorded pre- and posts-vibration testing. The inspected surfaces on the optic cell can be seen below in *Figure 20*.

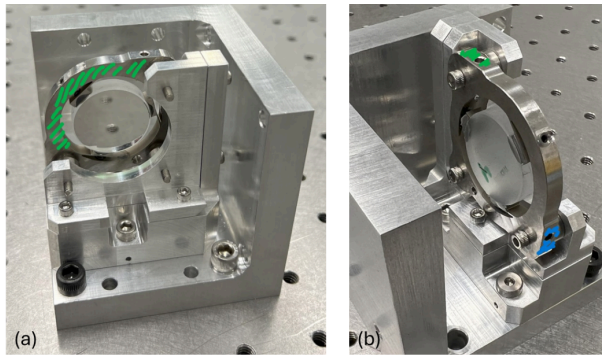

*Figure 20: (a) Inspected Surface Plane of the Cell (b) Inspected Point Circle of Location Holes*

For the optic stage, the vertical surfaces parallel to the YZ plane, fastener plane surfaces, and vertical surfaces parallel to the XZ plane were inspected. These inspected surfaces can be seen below in *Figure 21*

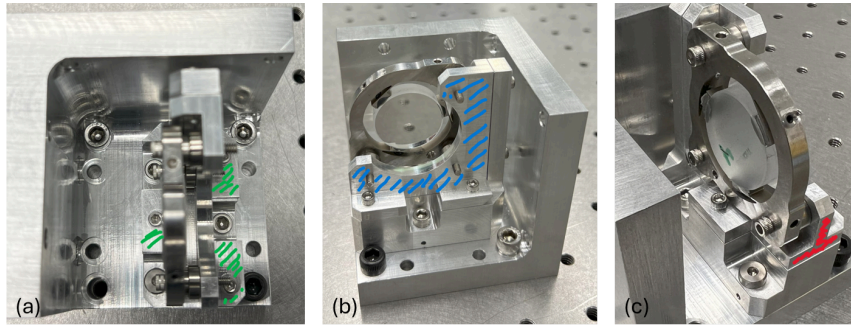

*Figure 21: (a) Inspected Fastener Plane Surface (b) Inspected Vertical Surface Parallel to XZ (c) Inspected Vertical Surface Parallel to YZ Plane*

The decenter stage inspections included the vertical surfaces parallel to YZ and XZ planes and a horizontal surface parallel to the XY plane. This configuration is presented below in *Figure 22*.

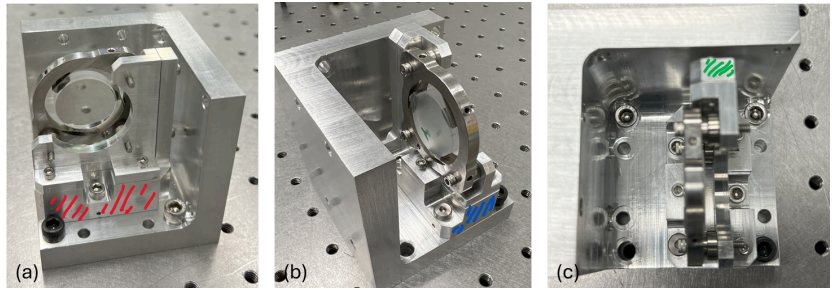

*Figure 22: (a) Inspected Vertical Surface Parallel to XZ (b) Inspected Vertical Surface Parallel to YZ (c) Inspected Horizontal Surface Parallel to XY Plane*

For thermal testing, a new coordinated system was created using the CAM2 software and measurement data file from the post-vibration evaluation. The measurements prior to thermal cycling were conducted post-vibration testing and were relevant to the vibration fixture setup. The reference coordinate system that was created can be seen below in *Figure 23*.

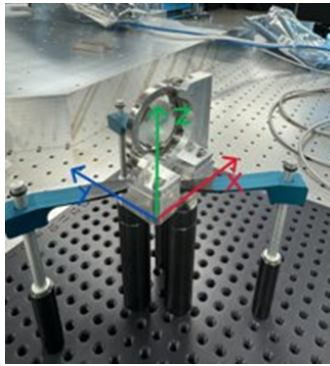

*Figure 23: Post-Thermal Testing Coordinate System Reference*

*Figure 24* depicts the surfaces and points that were evaluated to measure the optic cell, optic stage, and optic stage decenter, post-thermal testing.

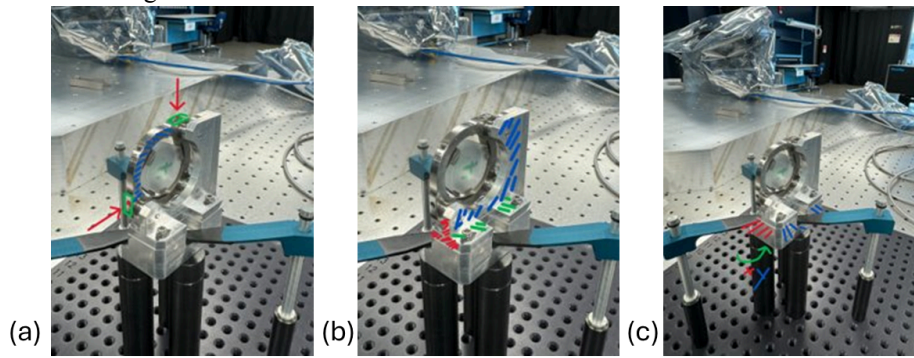

*Figure 24: (a) Surfaces and points measured for the optic cell evaluation (b) Surfaces measured for the optic stage evaluation (c) Surfaces measured for the optic stage decenter evaluation*

To find the angle and positional differences between pre- and post-vibration and thermal testing, MATLAB was used. With the assistance of MATLAB, statistical analysis was performed to identify any significant deviations. By using position differentials, the decentering and shifting of components were evaluated to determine if any significant deviations exceeded the defined uncertainty threshold. The uncertainty threshold utilized was based on FARO's stated MEP and the derived probe compensation error. Measurement uncertainty was considered to be 0.00078 inches.

# 4. RESULTS AND DISCUSSION

## 4.1 Optic Cell Assembly - Dynamic

The sine-sweep response for UUT1 in the x-axis is shown in *Figure 25a* and *25b*. Sine-Sweeps were performed similarly for each individual optic cell assembly in all three (3) orthogonal axes throughout testing. It was noted that all of the test articles exhibited nearly the same behavior with no major modes being activated, up to 3500 Hz. This observed lack of resonance can be attributed to the radial design of the cell which is inherently stiff. ISO 19683 and GEVS standards limit their dynamic spectra testing up to 2kHz, beyond which they transition to shock spectra. The stiffness of the radial design effectively suppresses any resonant frequencies that would otherwise be found within the dynamic spectra. This emphasizes the robustness of the assembly's ability to withstand lower-frequency mechanical events.

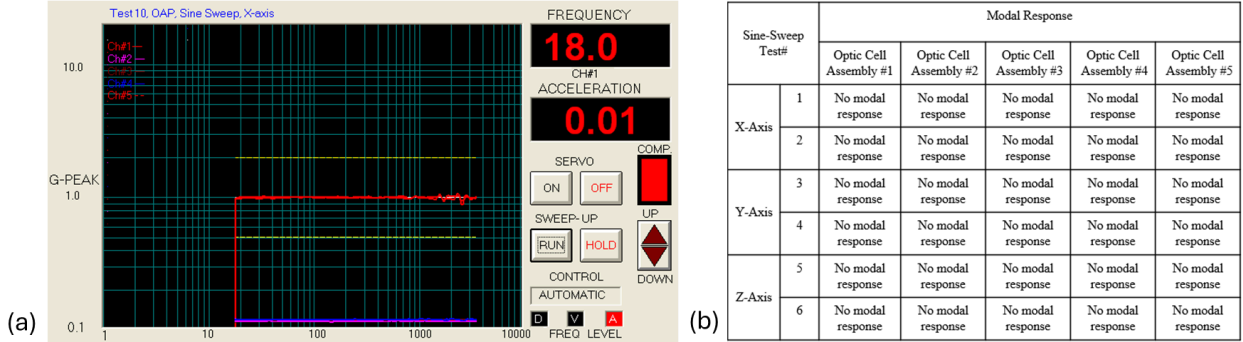

*Figure 25: (a) Sine-Sweep Optic Cell Assembly Response - UUT1 (b) Sine-Sweep Modal Response of Optic Cell Assembly*

Random vibration testing was performed in accordance with ISO19683, however the implemented test differed slightly as shown in *Figure 26a*. This slight modification to the frequency breakpoint allowed additional operating margin to the vibration shaker being used that had a limit load of 66 Grms. The random vibration dynamic response is shown in *Figure 26b*. As shown, UUT2 had the ISO19683 acceptance PSD applied to the test article similar to UUT1, UUT3 and UUT4. UUT5 had the ISO19683 qualification PSD applied to the test article. From *Figure 27a*, it is observed that the UUT2 test article, and other identified test articles, passed the acceptance test with no failure seen across the spectra which aligns with the sine-sweeps performed. From *Figure 27b*, it is observed that the UUT5 test article passed the qualification test with no failure seen across the spectra which also aligns with the prior sine-sweep test.

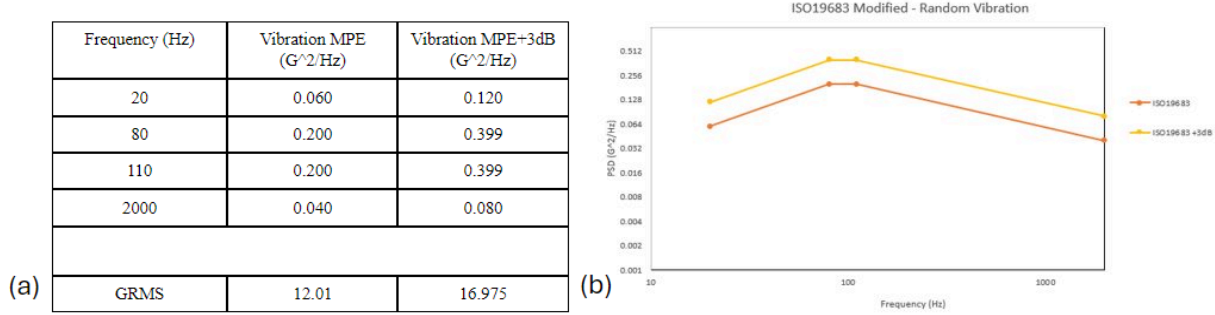

*Figure 26: (a) ISO19683 Modified - Random Vibration Breakpoint Table and (b) ISO19683 Modified - Random Vibration PSD Response*

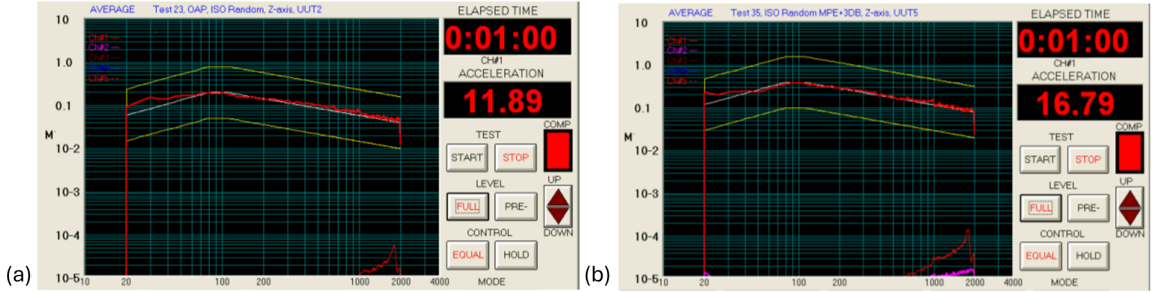

*Figure 27: (a) ISO19683 MPE Random Vibration Acceptance Response - UUT2 (b) ISO19683 MPE+3dB Random Vibration Qualification Response - UUT5*

## 4.2 Optic Cell Assembly - Mechanical Shock

The mechanical shock response for the 1000G target is shown in *Figure 28* with measured values seen in *Figure 29a*. As shown, a Tektronix TDS2001C Oscilloscope was used to measure and compare the responses of the control and UUT accelerometers. Based on the voltage response of the accelerometer, the shock (G) load could be extracted. Channel 1, the shock plate control, had a sensitivity of 10.41 mV/G and channel 2, the test article, had a sensitivity of 10.50 mV/G. During testing it was observed that the accelerometers that were being utilized, Dytran 3032A, had a reduced accuracy above the 1500G level, therefore for shock values that exceeded 1500G's, a linear interpolation of the drop data was utilized for identifying the acceptable drop height to achieve the specified target shock level as shown in *Figure 29b*. The final test, sequence number 14, was the maximum limit of the slide rail. The correlation between the target shock responses and measured shock responses instilled high confidence that the shock levels were satisfied.

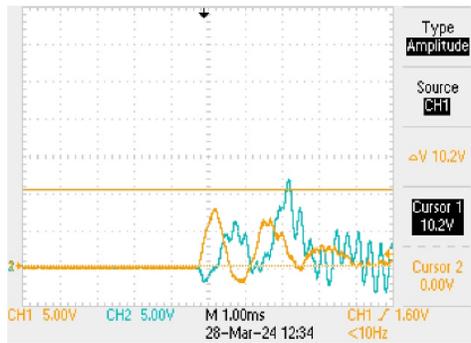

*Figure 28: Mechanical Shock Test Comparison - 1000G Target*

| Test Sequence Number | Target Shock Level (G) | Measured Shock Level (G) |
|---|---|---|
| 1 | 500 | 730 |
| 2 | 1000 | 1090 |
| 3 | 1500 | 1420 |
| 4 | 2000 | 2210 |
| 5 | 2500 | 2520 |
| 6 | 3000 | 3000 |
| 7 | 4000 | Extrapolated |
| 8 | 5000 | Extrapolated |
| 9 | 6000 | Extrapolated |
| 10 | 7000 | Extrapolated |
| 11 | 8000 | Extrapolated |
| 12 | 9000 | Extrapolated |
| 13 | 10000 | Extrapolated |
| 14 | 10650 | Extrapolated |

(a)

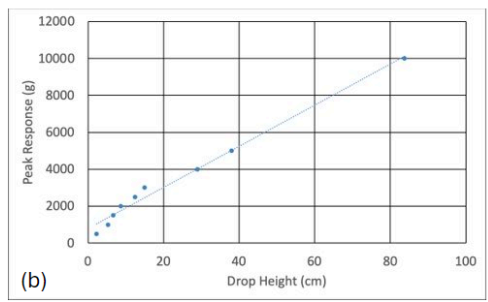

(b)

*Figure 29: (a) Mechanical shock response target shock and measured shock levels and (b) mechanical shock extrapolated peak responses in excess of 1500G*

Between each test sequence, the UUT was visually inspected for hardware failures (adhesive cracks or detachments, glass cracking or chipping, and optic cell deformations) with a Bausch and Lomb 10.5x to 45x stereo-microscope. No visual deformations or failures were observed to the UUT.

### 4.3 Optic Cell Assembly and Optic Assembly - Thermal Endurance

During the thermal endurance testing, four (4) cycles over the course of 25.33 hrs, the temperature profile was defined to have a ramp rate of 2°/min and a soak time of three (3) hrs at each extreme. As seen in *Figure 30*, the obtained thermocouple readings confirmed adherence to the defined profile. The temperature profile was followed closely with only slight thermal drift observed towards the end of the cycles. The thermal drift can likely be attributed to thermal gradients within the oven environment as well as slight variations in the conduction and insulation properties of the mounting surfaces. The thermal drift remained negligible throughout the duration of thermal cycling. Following the completion of thermal cycling, the optic cell assembly and optic assembly were visually inspected using a high-resolution microscope, upon which no visible signs of optic displacement or adhesive failure were discovered. The optic assembly underwent further post-testing inspection using a Faro Arm.

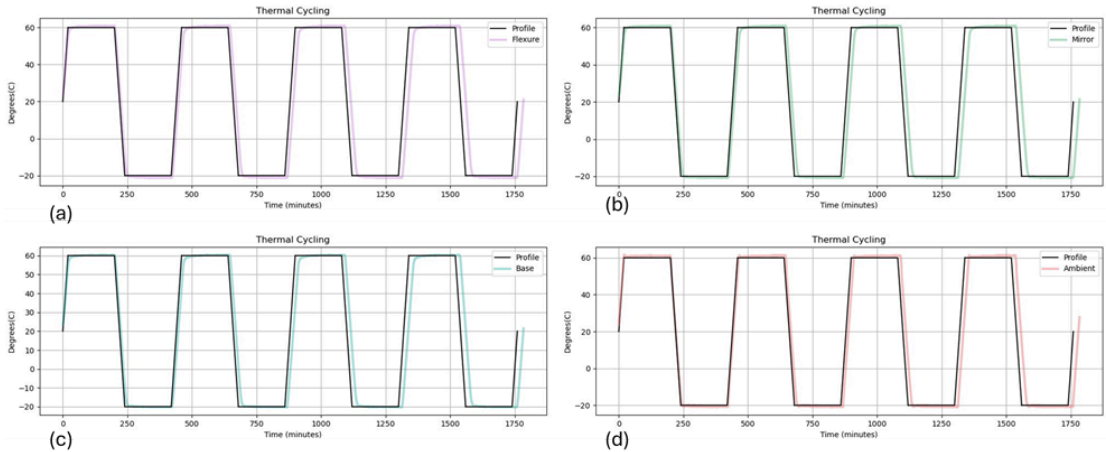

*Figure 30: (a) Flexure thermocouple reading (b) Mirror thermocouple reading (c) Base thermocouple reading (d) Ambient thermocouple reading*

### 4.4 Optic Assembly-Dynamic

The optic assembly underwent sine-sweep tests before and after the random vibration tests to confirm that the test unit experienced no premature failure. Sine-sweeps were performed in all three (3) orthogonal axes. The sine response of the optic assembly can be seen below in *Figure 31*. The sine-sweep analysis identifies the first significant mode around 1200Hz, which aligns with the expected mode as predicted by ANSYS simulations. The minimal discrepancy between the analytical predictions and experimental responses confirms the model is well-correlated with the actual build and definition of the optic assembly.

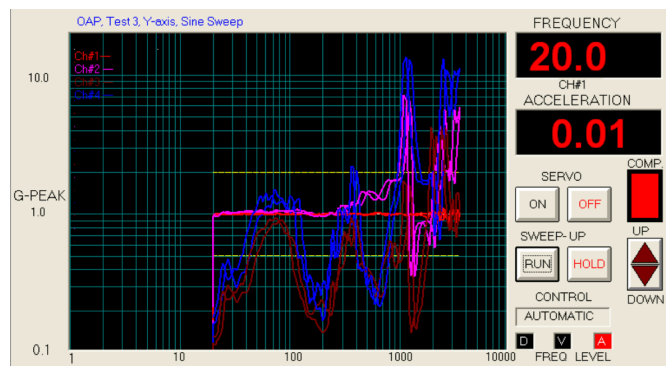

*Figure 31: Sine-Sweep Optic Assembly Response*

For random vibration testing, due to the increased mass of the test article and the limitations of the shaker table, the GEVS PSD was pursued. To qualify the assembly, the assembly was tested to GEVS MPE random vibration acceptance levels seen in *Figure 32*. The results seen in *Figure 33* confirm that the UUT passed the acceptance test, with no failures observed across the spectra. This outcome aligns with the results of the sine-sweep test, further validating the robustness of the optic assembly.

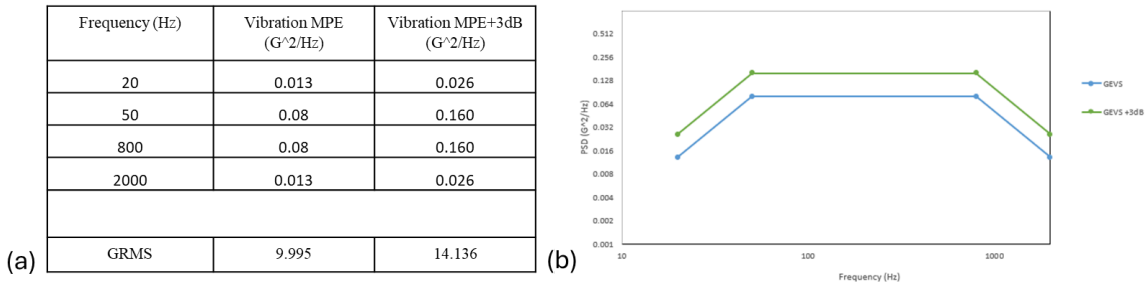

| Frequency (Hz) | Vibration MPE (G^2/Hz) | Vibration MPE+3dB (G^2/Hz) |
|---|---|---|
| 20 | 0.013 | 0.026 |
| 50 | 0.08 | 0.160 |
| 800 | 0.08 | 0.160 |
| 2000 | 0.013 | 0.026 |
|  |  |  |
| GRMS | 9.995 | 14.136 |

(a) (b)

*Figure 32: (a) GEVS - Random Vibration Breakpoint Table and (b) GEVS - Random Vibration PSD Response*

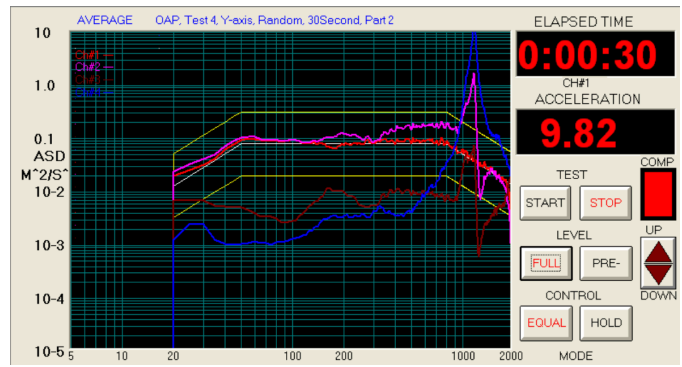

*Figure 33: GEVS Random Vibration Qualification Response - Optic Assembly*

### 4.5 Post-Testing Analysis

### 4.5.1 Interferometric Analysis

Interferometric analysis was conducted with an 85% clear aperture and 100% clear aperture to assess the wavefront error of the optical mounts post vibration testing. To determine the wavefront error induced by the structural tests, the interferometric data was analyzed post testing with the pre-testing data as a base line. The comparison of the pre- and post- wavefront data allowed for the impacts of the structural loads on the optical mounts to be quantified. Due to resource constraints, only one mount was interferometrically tested prior to the vibe test. As such, the original pre-vibe frame was reused as the reference baseline for all subsequent post-vibe analyses. The post-vibe interferometry of the mounts can be seen below in *Figure 34*.

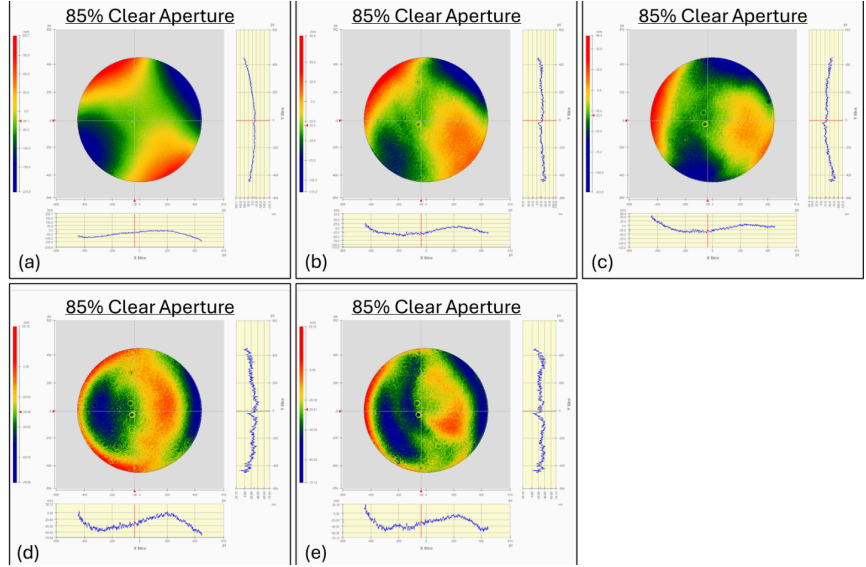

*Figure 34: Post-Vibration Interferometric Analysis (a) UUT1 (b) UUT2 (c) UUT3 (d) UUT4 (e) UUT5*

The post-vibration interferometric analysis of the optical mounts (UUT-UUT5) indicates the wavefront error profiles indicative of their structural resilience. Among the units tested, UUT 1 performed the worst with significant defocus and optical aberrations, resulting in peak-to-valley (PV) wavefront error of 420nm and root-mean-square (RMS) value of 74nm. The best performers were the radial designs, UUT4 and UUT5 which can be seen in *Figure 34d* and *Figure 34e*. Both had low PV wavefront error and outperformed their pre-vibration baseline, proving their resilience in maintaining their structural and optical integrity under vibrational stress. Despite the observed variations, all units still met their optical specifications. The respective values for all UUT's be seen below in *Table 5*.

| Post Vibe | UUT1 | UUT2 | UUT3 | UUT4 | UUT5 |
|---|---|---|---|---|---|
| PV [nm] | 420 | 199 | 190 | 103 | 104 |
| RMS [nm] | 74 | 27 | 24 | 14 | 13 |

*Table 5: Post-Vibration Flexure Value Comparison Table*

### 4.5.2 Faro Arm Analysis

The optic cell and assembly were analyzed pre- and post- vibration testing and thermal testing to determine the impact of the mechanical stresses on the optical alignment. *Table 6* shows the error budget allowance utilized during testing and analysis. The initial error budget allowance allowed for 10 um of x, y, z perturbation and 0.001° rotation about x, y, z for optic misalignment. Based on the Faro Arm reported MPE, and user derived probe compensation error, measurement uncertainty was considered to be 0.00078 inches.

| Misaligning / solid body motion | | Surface irregularity | E2 |
|---|---|---|---|
| All optics perturbation | +-(um, deg) | All optics | |
| x | 10 | 5 nm RMS Sag error using Zernike 5-100 | |
| y | 10 | | |
| z | 10 | | |
| Rx | 0.001 | | |
| Ry | 0.001 | | |
| Rz | 0.001 | | |

*Table 6: Error Budget*

*Table 7a* shows the pre- and post- vibration test measurements for the positioning of the optic cell. *Table 7b* shows the distance between the measured point and the boundary of the uncertainty interval. The positional changes for the

Top and Side points fall within this uncertainty range. The change in the Y-distance of the cell plane to the origin exceeds the uncertainty threshold, with a deviation of 0.00032" beyond the boundary of the confidence interval given by +/- 0.00078".

| | Direction | Top point (in) | Side Point (in) | Y Dist. To Cell Plane (in) |
|---|---|---|---|---|
| Pre Vibe | | | | |
| | x | 1.6433 | 2.9457 | |
| | y | 1.7335 | 1.7370 | 1.8324 |
| | z | 2.6758 | 1.3700 | |
| Post Vibe | | | | |
| | x | 1.6431 | 2.9454 | |
| | y | 1.7330 | 1.7370 | 1.8313 |
| | z | 2.6760 | 1.3705 | |
| Delta | | | | |
| | x | 0.0002 | 0.0003 | |
| | y | 0.0005 | 0.0000 | 0.0011 |
| | z | -0.0002 | -0.0005 | |

(a)

| | Plane | Angle off parallel - Cell plane to XZ plane at origin (Deg) |
|---|---|---|
| Pre Vibe | | |
| | XY | 0.5680 |
| | XZ | 0 |
| | YZ | 0.3854 |
| | 3D Space | 0.6864 |
| Post Vibe | | |
| | XY | 0.5579 |
| | XZ | 0 |
| | YZ | 0.3727 |
| | 3D Space | 0.6709 |
| Delta | | |
| | XY | 0.0101 |
| | XZ | 0.0000 |
| | YZ | 0.0127 |
| | 3D Space | 0.0155 |

(b)

*Table 7: (a) Measurements of Optic Cell Position Pre- and Post- Vibration Testing (b) Angle of Rotation of Optic Cell Pre- and Post- Vibration Testing*

*Table 8* shows the pre- and post- vibration test measurements for the positioning of the optic stage. The data captures the positional changes across all planes (x, y, z) and the angle of rotation from the test planes. The observed changes in position and rotation are all within the established measurement uncertainty of 0.00078".

| | Direction | Position - Fastener Plane (in) | Position - Vertical Surface Parallel to XZ plane (in) | Position - Vert Surface Parallel to YZ Plane (in) |
|---|---|---|---|---|
| Pre Vibe | | | | |
| | x | 0 | 0 | 2.9827 |
| | y | 0 | 2.2118 | 0 |
| | z | 0.8947 | 0 | 0 |
| Post Vibe | | | | |
| | x | 0 | 0 | 2.9827 |
| | y | 0 | 2.2113 | 0 |
| | z | 0.8950 | 0 | 0 |
| Delta | | | | |
| | x | 0.0000 | 0 | 0 |
| | y | 0.0000 | 0.0005 | 0 |
| | z | -0.0003 | 0 | 0 |

(a)

| | Plane | Angle of Rotation - Description (Deg) |
|---|---|---|
| Pre Vibe | | |
| | XY | 0.2885 |
| | XZ | 0 |
| | YZ | 0.3260 |
| | 3D Space | 0.4353 |
| Post Vibe | | |
| | XY | 0.2882 |
| | XZ | 0 |
| | YZ | 0.3239 |
| | 3D Space | 0.4335 |
| Delta | | |
| | XY | 0.0003 |
| | XZ | 0.0000 |
| | YZ | 0.0021 |
| | 3D Space | 0.0018 |

(b)

*Table 8: (a) Measurements of Optic Stage Position Pre- and Post- Vibration Testing (b) Angle of Rotation of Optic Stage Pre- and Post- Vibration Testing*

*Table 9* shows the pre- and post- vibration test measurements for the positioning of the optic decenter stage. The data captures the positional changes across all planes (x, y, z) and the angle of rotation from the test planes. The measurement shows the movement in the Y-direction of the vertical XZ surface plane is greater than that of the uncertainty with a deviation of 0.00052" beyond the boundary of the confidence interval given by +/- 0.00078".

| | Direction | Position - Vert Surface Parallel to XZ (in) | Position - Vertical Surface Parallel to YZ (in) | Position - Horizontal surface Parallell to XY (in) |
|---|---|---|---|---|
| Pre Vibe | | | | |
| | x | 0 | 2.9964 | 0 |
| | y | 2.5552 | 0 | 0 |
| | z | 0 | 0 | 2.7055 |
| Post Vibe | | | | |
| | x | 0 | 2.9964 | 0 |
| | y | 2.5539 | 0 | 0 |
| | z | 0 | 0 | 2.7054 |
| Delta | | | | |
| | x | 0.0000 | 0 | 0 |
| | y | 0.0013 | 0.0000 | 0 |
| | z | 0.0000 | 0 | 0.00010 |

(a)

| | Plane | Angle of Rotation - Description (Deg) |
|---|---|---|
| Pre Vibe | | |
| | XY | 0.1019 |
| | XZ | 0 |
| | YZ | 0.0634 |
| | 3D Space | 0.1200 |
| Post Vibe | | |
| | XY | 0.1046 |
| | XZ | 0 |
| | YZ | 0.0587 |
| | 3D Space | 0.1200 |
| Delta | | |
| | XY | -0.0027 |
| | XZ | 0 |
| | YZ | 0.0047 |
| | 3D Space | 0.0000 |

(b)

*Table 9: (a) Measurements of Optic Decenter Stage Position Pre- and Post- Vibration Testing (b) Angle of Rotation of Optic Decenter Stage Pre- and Post- Vibration Testing*

The movement within the assembly due to the vibration tests was small and likely at the resolution limit of the Faro arm. The deviation beyond the boundary of confidence can likely be attributed to the inherent measurement uncertainty or probe compensation errors.

The Faro Arm analysis of the optic sub-assembly post-thermal testing utilized the same error budget as seen in *Table 6*, as well as the same user-derived uncertainty. The surfaces and points measured for the post-thermal test are those seen in *Figure 24* in the Methodology section. *Table 10* shows the pre- and post- thermal test measurements for the optic cell position. Both the top and side location hole measurements exhibit a shift in the z-direction, indicating a potential downward translation of the cell in the Z-direction of ~0.0012'' as well as a possible small rotation around the X and Z-axis.

| Inspection | \multicolumn{5}{c}{Measurements of Optic Cell to Reference Frame Pre- and Post-Thermal Test} | |
|---|---|---|---|---|---|---|
| | Direction | Top point (in) | Side Point (in) | Y Dist. To Cell Plane (in) PointOnPlane | Reference | Angle off parallel - Cell plane to XZ plane at origin (Deg) |
| Pre-Thermal Test | | | | | | |
| | x | 1.3523 | 0.0500 | | XY | 0.6625 |
| | y | 0.8210 | 0.8146 | 0.7212 | XZ | 0.0000 |
| | z | 2.6760 | 1.3705 | | YZ | 0.3727 |
| | | | | | 3D Space | 0.7601 |
| | Flatness | | | 0.0006 | | |
| Post Thermal Test | | | | | | |
| | x | 1.3523 | 0.0498 | | XY | 0.6546 |
| | y | 0.8204 | 0.8143 | 0.7212 | XZ | 0.0000 |
| | z | 2.6748 | 1.3693 | | YZ | 0.3810 |
| | | | | | 3D Space | 0.7574 |
| | Flatness | | | 0.0003 | | |
| Delta | | | | | | |
| | x | 0.0000 | 0.0002 | | XY | 0.0079 |
| | y | 0.0006 | 0.0003 | 0.0000 | XZ | 0.0000 |
| | z | 0.0012 | 0.0012 | | YZ | -0.0083 |
| | | | | | 3D Space | 0.0027 |
| | Flatness | | | 0.0003 | | |

*Table 10: Measurements of Optic Cell Position Pre- and Post- Thermal Testing*

*Table 11* shows the pre- and post- thermal test measurements for the optic stage position. The fastener plane exhibited relatively poor flatness, which can be attributed to machining imperfections or the presence of small amounts of staking material. The measurements indicate a potential rotation, primarily concentrated around the X-axis. Additionally, the vertical surface measurements show a possible movement in the y-direction, along with a relatively large rotation around the Z-axis.

| | Direction | Position - Fastener Plane (in) | Reference | Angle of Rotation - Fastener Plane (Deg) | Direction2 | Position - Vertical Surface parallel to XZ plane (in) | Reference2 | Angle of Rotation - Vertical Surface P to XZ (Deg) |
|---|---|---|---|---|---|---|---|---|
| **Pre-Thermal Test** | | | | | | | | |
| | x | 0.0000 | XY | 0.0000 | x | 0.0000 | XY | 0.6625 |
| | y | 0.0000 | XZ | 0.1795 | y | 0.3425 | XZ | 0.0000 |
| | z | 0.8950 | YZ | 0.2018 | z | 0.0000 | YZ | 0.3727 |
| | | | 3D Space | 0.2700 | | | 3D Space | 0.5091 |
| | Flatness | 0.0021 | | | Flatness | 0.0006 | | |
| **Post Thermal Test** | | | | | | | | |
| | x | 0.0000 | XY | 0.0000 | x | 0.0000 | XY | 0.3658 |
| | y | 0.0000 | XZ | 0.1731 | y | 0.3437 | XZ | 0.0000 |
| | z | 0.8946 | YZ | 0.1495 | z | 0.0000 | YZ | 0.3230 |
| | | | 3D Space | 0.2287 | | | 3D Space | 0.4880 |
| | Flatness | 0.0019 | | | Flatness | 0.0005 | | |
| **Delta** | | | | | | | | |
| | x | 0.0000 | XY | 0.0000 | x | 0.0000 | XY | 0.2967 |
| | y | 0.0000 | XZ | 0.0064 | y | -0.0012 | XZ | 0.0000 |
| | z | 0.0004 | YZ | 0.0523 | z | 0.0000 | YZ | 0.0497 |
| | | | 3D Space | 0.0413 | | | 3D Space | 0.0211 |
| | Flatness | 0.0002 | | | Flatness | 0.0001 | | |

Table 11: Measurements of Optic Stage Position Pre- and Post-Thermal Testing

*Table 12* shows the measurements of the decenter stage position post-thermal testing. The decenter stage was used to set up the coordinate system for pre/post thermal test measurements. The pre-thermal measurements of the coordinate system XY plane were conducted by surface-touching the bottom of the assembly. This may explain the 0.0012" downward shift in the Z-direction, which was consistent with the shifts noted in the optic cell location points.

| | Direction | Position - Vert Surface Parallel to XZ (in) | Reference | Angle of Rotation - Vert Surface Parallel to XZ (Deg) | Direction 2 | Position - Vertical Surface Parallel to YZ (in) | Reference2 | Angle of Rotation - Vertical Surface Parallel to YZ (Deg) | Direction3 | Position - Horizontal surface Parallel to XY (in) | Reference3 | Angle of Rotation - Horizontal Surface Parallel to XY (Deg) |
|---|---|---|---|---|---|---|---|---|---|---|---|---|
| **Pre-Thermal Test** | | | | | | | | | | | | |
| | x | 0.0000 | XY | 0 | x | -0.0005 | XY | 0.0301 | x | 0.0000 | XY | 0.0000 |
| | y | -0.0004 | XZ | 0 | y | 0.0000 | XZ | 0.0324 | y | 0.0000 | XZ | 0.2584 |
| | z | 0.0000 | YZ | 0.0587 | z | 0.0000 | YZ | 0.00000 | z | 2.7054 | YZ | 0.0992 |
| | | | 3D Space | 0.0587 | | | 3D Space | 0.0442 | | | 3D Space | 0.2768 |
| | Flatness | 0.0003 | | | Flatness | 0.0004 | | | Flatness | 0.0002 | | |
| **Post Thermal Test** | | | | | | | | | | | | |
| | x | 0 | XY | 0.0074 | x | -0.0002 | XY | 0.0000 | x | 0.0000 | XY | 0.0000 |
| | y | -0.0005 | XZ | 0 | y | 0.0000 | XZ | 0.0407 | y | 0.0000 | XZ | 0.2510 |
| | z | 0 | YZ | 0.0617 | z | 0.0000 | YZ | 0.0000 | z | 2.7042 | YZ | 0.1044 |
| | | | 3D Space | 0.0622 | | | 3D Space | 0.0407 | | | 3D Space | 0.2710 |
| | Flatness | 0.0009 | | | Flatness | 0.0003 | | | Flatness | 0.0002 | | |
| **Delta** | | | | | | | | | | | | |
| | x | 0.0000 | XY | -0.0074 | x | -0.0003 | XY | 0.0301 | x | 0.0000 | XY | 0.0000 |
| | y | 0.0001 | XZ | 0 | y | 0.0000 | XZ | -0.0083 | y | 0.0000 | XZ | 0.0074 |
| | z | 0.0000 | YZ | -0.0030 | z | 0.0000 | YZ | 0.0000 | z | 0.0012 | YZ | -0.0052 |
| | | | 3D Space | -0.0035 | | | 3D Space | 0.0035 | | | 3D Space | 0.0058 |
| | Flatness | -0.0006 | | | Flatness | 0.0001 | | | Flatness | 0 | | |

Table 12: Measurements of Optic Decenter Stage Position Pre- and Post-Thermal Testing

Considerable time passed between the thermal test and the post-test measurements. The measurement setup prior to the thermal test also differed from the setup post-thermal test. Differences in the alignment of the Faro Arm, the positioning of the assembly, and the method of securing the components, all could have introduced inconsistencies. The two observed positional measurement changes that exceeded the MPE defined by the Faro arm, may be attributed to the differences in the measurement setup. This setup variation could also have contributed to the rotational shift seen about the X-axis of the optic stage. The most substantial rotational changes were found around the Z-axis and X-axis. These changes were found to be two (2) to three (3) times the order of magnitude larger than the original error budget values for rotation of optics in the assembly. This suggests that the differences in the pre- and post-test measurements processes had a significant impact on the accuracy of the rotational data, leading to larger than expected deviations from the initial error budget.

# 5. CONCLUSION

Our research demonstrates the successful development and validation of precise, adjustable, flexure mounts that are capable of supporting small optics in space applications. The mounts, designed to support small optics up to 50mm in diameter and over 100 grams, have shown resilience through comprehensive thermal testing, multi-axis vibration testing, and mechanical shock testing in accordance with ISO 19683 standards.

The results indicate that our mounts maintain wavefront quality within lambda/10, ensuring minimal wavefront degradation after undergoing harsh environmental conditions. The radial design effectively suppresses resonant frequencies up to 3500 Hz, and demonstrates the capability to withstand mechanical shock testing and thermal testing with no optic displacement or adhesive failure. Interferometric and Faro Arm analyses confirmed that the mounts still met the stringent positional and wavefront quality requirements with minor deviations well within the acceptable limits.

This advancement in mounting technology holds significant promise for offering a reliable and efficient solution to reduce development time and cost for high-precision optical systems. By testing to ISO +3dB, the mounts do not require requalification for various launch vehicles which further streamlines the integration process. Our work thus helps to further contribute to the ongoing advancement in the design and implementation of optical mounts for astronomy, laser communication, and other critical space-based applications.

## ACKNOWLEDGMENTS

Portions of this research were supported by funding from the Technology Research Initiative Fund (TRIF) of the Arizona Board of Regents and by generous anonymous philanthropic donations to the Steward Observatory of the College of Science at the University of Arizona.